# Next Challenges in Bringing Artificial Immune Systems to Production in Network Security

Michael Hilker
University of Luxembourg
6, Rue Richard Coudenhove-Kalergi
1359 Luxembourg
michael.hilker@uni.lu

## ABSTRACT

The human immune system protects the human body against various pathogens like e.g. biological viruses and bacteria. Artificial immune systems reuse the architecture, organization, and workflows of the human immune system for various problems in computer science. In the network security, the artificial immune system is used to secure a network and its nodes against intrusions like viruses, worms, and trojans. However, these approaches are far away from production where they are academic proof-of-concept implementations or use only a small part to protect against a certain intrusion. This article discusses the required steps to bring artificial immune systems into production in the network security domain. It furthermore figures out the challenges and provides the description and results of the prototype of an artificial immune system, which is SANA called.

## Categories and Subject Descriptors

I.2 [**Artificial Intelligence**]: Distributed Artificial Intelligence; C.2.4 [**Distributed Systems**]: Distributed Applications—*deployment, administration, self-management*

## 1. INTRODUCTION

Network security is the domain where a network is protected against intrusions. These intrusions are automatic attacks like viruses, worms, and trojans as well as manual attacks performed by hackers or normal users trying to gain access to resources where they normally have no access. Different various protection systems try to protect the network and its nodes where the protection system consists of several protection components. The protection components are a software or hardware solution using a specific type of tasks for detecting a set of intrusions. Examples are the host-based antivirus softwares, firewalls, and maleware guards or the network-based packet filters and intrusion detection systems (IDS) [21]. These components run different tasks like observing the file access and system calls on a node or header and content scanning of packets. The protection system is mostly a collection of protection components lacking of collaborative work and sophisticated information management in order to identify upcoming more and more intelligent and adaptive intrusions. The systems do not check itself or use the information in order to identify infected nodes - nodes with a running intrusion -, not proper working components, and abnormal behavior.

Novel approaches are the artificial immune systems. These systems reuse the architecture, organization, and workflows of the human immune system for various domains in computer science where this article focusses on the network security domain. However, the current approaches in artificial immune system are far away from productive work and mainly implement a few parts of the human immune system as an academic proof-of-concept implementation, which focus on detecting a certain intrusion. The steps to bring an artificial immune system into production so that the security systems profit from the advantages are discussed in this article. Afterwards, the prototype SANA of an artificial immune system is introduced and its results in different attack scenarios are discussed.

## 2. THE NETWORK SECURITY DOMAIN

In the network security domain is the goal to identify the intrusions using protection systems; a great overview about network security is in the book [24]. The protection systems consist normally of the following protection components:

**Antivirus Software** The antivirus software observes a node whether an infected file is accessed or a suspicious system call occurs. The intrusions are mostly described using a signature where pattern matching is used. When an intrusion is found asks the antivirus software the user how to proceed.

**Firewall** The network traffic consists of packets where the firewall analyses the packet header in order to find intrusions. When an intrusion is found is the user consulted for further steps.

**Packet Filter** In the network equipment is the packet filter installed. It analyses the packet header for intrusions. Furthermore, it defines a network security policy with allowed and disallowed traffic, i.e. allowed and disallowed ports and network protocols.

**Intrusion Detection System** Important nodes - e.g. Internet gateway and email server - require additional support because they are more likely an aim for attackers than normal nodes. Therefore, intrusion detection systems are used, which check each packet completely, observe the node for suspicious behavior, and report warnings and alerts directly to the administrator [3, 5, 21].

**Other Systems** Different other protection systems exist, which perform certain tasks for network security. Examples are virus throttles slowing down the propagation of viruses [27] or automatic analyzing systems of infected nodes [20].

The protection system defines the used protection components, the configuration, and the workflows. Additionally, it defines in which way the administrator maintains the network, keeps the system up-to-date, and the response workflows when an intrusion is identified.

Current protection systems use a fully centralized approach. Each node of the network has one or more client softwares, which are administrated using a single management server - client-server architecture. The administrator installs and configures in each node antivirus software and firewall, in the network equipment the packet filters, and in important nodes the IDS. The client softwares are the already described components and observe the node for suspicious behavior. The alerts - e.g. identification of an intrusion - are sent to the user of the node, who decides how to proceed. The warnings are sent to the management server and are manually evaluated by the administrator. The management server coordinates the client software in providing updates and administrative tasks. It also checks the nodes whether the client software runs or not. The different client softwares do not collaborate as well as the messages from different nodes are not combined evaluated in order to identify abnormal behavior. Furthermore, the system does not check itself for the identification of not proper working or outdated components and it has serious problems with infected nodes.

For the evaluation of protection system exist different criteria [5, 19]:

**Completeness** The protection system should secure the nodes of the network. Furthermore, it should secure all nodes against all known intrusions.

**Production Tolerant** The production in the network should not be influenced by the security system - reducing of the false-positives.

**Efficiency** The resources should be used efficiently so that the protection system does not require too many resources. Furthermore is this important to solve the packet loss problem where IDS stop checking packets when a certain load is reached [22].

**Easy Usage** The security system should use as many possible automatic workflows so that the administration is reduced.

**Self-Checking** The system should check itself regularly in order to identify infected nodes, not proper working or outdated security components, and abnormal behavior in the network.

**Adaptively** In order to identify the current intrusions as well as modified and novel intrusions, the security system should adapt to the current situation and provide adaptive workflows to identify novel intrusions.

**Coping with upcoming Intrusions** Novel intrusions analyze the protection system and use weak points, camouflage itself so that it cannot be detected anymore, and social engineering is more and more used where the normal users are mislead to provide internal information as e.g. passwords.

**Implementation, Maintenance, Updates, Extension** These points should be simplified so that the administrator is able to introduce novel techniques and updates into the system quickly. The system should check and repair itself, and the implementation should be fast. In addition, the system should provide a status snapshot when the administrator demands it.

In the network security domain, common used protection systems use a centralized approach using the client-server architecture. The clients demand information from the server and the clients perform the tasks with a reporting to the server. Current protection systems have serious problems in each of these criteria and in coping with upcoming more and more intelligent and adaptive intrusions. Examples of these intrusions are the theoretical bradley virus [7] and metamorphic or polymorphic viruses that change its signature in every propagation [25]. The emerges out of the static architecture with standard pattern matching algorithm. Thus novel approaches should dynamically adapt to the current situation and perform different analysis and combine the results and information to identify novel intrusions. Existing artificial immune systems for the networks security domain are discussed in the next section.

## 3. ARTIFICIAL IMMUNE SYSTEMS FOR NETWORK SECURITY

The artificial immune system is a modeling of the human immune system for a specific application domain; in this article the domain is network security. Details about the human immune system are not explained in this article and it is referred e.g. to [16]. In the design and implementation of an artificial immune system, mostly a few parts of the human immune system are modeled. The article [8] defines and evaluates four different methods approaching from the research in artificial immune systems [6]:

**Artificial Negative and Positive Selection** The artificial cells are generated randomly and are afterwards evaluated whether they are tolerant to normal network traffic or not - negative selection - and whether they detect abnormal network traffic or not - positive selection.

**Danger Theory** The idea is that the artificial cells release signals describing their status, e.g. safe signals and danger signals. The various artificial cells use the signals in order to adapt their behavior.

**Artificial Clonal Selection and Hypermutation** The artificial cells respond when an intrusion is found: it firstly copies itself heavily so that the number of this artificial cell increases and the intrusion is found in several nodes. Second, the artificial cell mutate in order to identify the intrusion more properly, e.g. in adapting the internal patterns for finding intrusions.

**Artificial Immune Networks** These networks describe a mathematical model of antibodies and antigens that bind and interfere each other. The system reacts to the current situation of bindings and its strength.

In [8] is stated that these four methods are the most promising as well as novel approaches of the artificial immune systems, which are significant different to existing approaches of computer science in general. However, the negative selection is not appropriate for anomaly detection - identification of intrusions according to behavior analysis [23]. The organization of the immune system is important and different to the current protection systems in network security: the immune cells work autonomously as a mobile entity. The workflows are fragmented in different small tasks, which are performed by different cells. The high number of cells ensures redundancies so that a partly breakdown does not influence the performance of the overall system. The lymph nodes are a meeting point between various cells and they respond to events in the network through releasing immune cells and antigens. The bone marrow and thymus releases continuously novel immune cells in order to keep the population up-to-date. The cell communication enables the collaboration between immune cells and the system manages itself so that the whole body is secured. The self-checking and -healing identifies and removes not proper working immune cells. This organization should be used when an artificial immune system is deployed so that the fault tolerance and redundancies are available.

For the network security domain exist different approaches to use an artificial immune system or algorithms motivated by the human immune system:

**ARTIS/LISYS** This artificial immune system secures a broadcast network against intrusions. The artificial cells reside in the nodes and check the network packets for certain patterns. A pattern is a string containing the information about the source- and destination IP and port as well as the used network protocol. The artificial cells are generated according to the human immune system: the cells are randomly generated and the appropriate cells are selected using the positive and negative selection. More details about this approach can be found e.g. in [15].

**LIBTISSUE** Aickelin and his team implemented this artificial immune system simulator in a client/server architecture [2, 26]. The data collector are distributed over the clients and the analyzing part is centralized in the server. Herein, e.g. the dentritic cells of the innate immune system are implemented [9, 10] and approaches of the danger model [1].

**CIDS** Dasgupta introduces approaches to use an artificial immune system in the network security domain [4]. The used architecture is a multi-agent system with roaming agents performing different tasks. However, the architecture is different to the human immune system.

**Other approaches** Other artificial immune systems for network security focus mostly on the multi-agent architecture without the biological-motivated architecture. In some approaches are e.g. a broadcast network used so that the artificial cells must not move or only the capturing of information is distributed and the analysis is centralized. An example is explained in [11] discussing the generation of immune algorithms for evolutionary detectors but the organization of the system is not discussed. In [17] is a blueprint of an artificial immune system described where several parts are similar in the SANA system introduced below. In [18] are several algorithms analyzed and a framework of an artificial immune system introduced.

The next section describes challenges in the process of bringing an artificial immune system into production in the network security domain, which also copes with the upcoming requirements due to future trends in intrusions.

## 4. NEXT DESIGNING AND IMPLEMENTATION TASKS

The architecture between artificial immune systems and common used protection systems is different. In contrast to the client-server architecture of common used protection systems, the artificial immune systems implement a distributed architecture where lightweighted, mobile, and autonomous working artificial cells perform the required tasks. In order to run these cells in each node, some kind of middleware must be installed on each node, which handles the access to the resources of the node and also includes common used security components. The middleware should also ensure that only allowed artificial cells can access the resources and solve other security issues. Furthermore, the middleware should contain the knowledge for the normal production so that the artificial cells are lightweighted and platform independent. The middleware should distinguish the operating and security system so that legal evidences can be saved and the system is checked from outside.

For the maintenance of the protection system are workflows required. The administrator demands regularly a status snapshot of the system as well as the administrator wants to know the current status of the system. Therefore, all nodes should collect status information and specialized nodes provide quickly a summary status snapshot. However, the system should also analyze the warnings and alerts autonomously so that the load of the administrator is reduced. The administrator should be able to access all nodes and all components. The implementation of the system should be feasible and fast where extensions and novel techniques should be quickly deployed to all components of the system. Updates - e.g. of the database of known intrusions - are frequent and the system should include the updates as fast as possible e.g. through positive and negative selection. The installation of updates and extensions should be monitored so that unsuccessful installations are detected and reported.

Infected nodes or not proper working components are a serious problem in current security systems. The artificial immune system should identify these components through self-checking workflows. Furthermore, it should develop a strategy how to disinfect the nodes or to repair the components - self-repairing and -healing. The information, which are gathered in this process, should be included in further protection processes - learning of the system.

Due to the enormous number of artificial cells and to enable novel workflows, the information management must be more sophisticated implemented. The different artificial cells perform only small tasks and several cells have to cooperate so that the goals are reached. Therefore, a communication protocol should be implemented where a cell informs a set of cells about a certain event - point to multi-point communication. Then, the artificial cells should use the communication protocol for collaborative work where e.g. the danger theory can be used and self management organizes the artificial cells. Another point is that information from different nodes and from different analysis processes are used in order to identify suspicious behavior beside the standard processes of network security.

A protection system should secure the network against all attacks where especially modified or novel attacks are hard to prevent. Current protection systems use mostly the signature based approach extended with some heuristics and, thus, have serious problems with novel attacks. Artificial immune systems should be adaptive in order to cope with the more and more intelligent and adaptive intrusions. Therefore, both the artificial cells as well as the overall system should adapt to the current situation. The adaptively in artificial cells can e.g. implemented using the artificial clonal selection and hypermutation as well as internal measurements.

These points are problems of protection systems and must be solved when an artificial immune system is deployed in the network security domain. SANA is a framework for an artificial immune system and is explained in the next section.

## 5. SANA - ARTIFICIAL IMMUNE SYSTEM

*SANA* is a framework for a distributed protection system where the architecture, design, and workflows are mostly biological motivated [14] and the performance is increased due to an enhancement of the organization of the protection components. The security system is an artificial immune system using artificial cells as well as common used protection components in order to secure a network against intrusions. The protection components, e.g. antivirus softwares, firewalls, packet filters, intrusion detection systems, and artificial cells, use various approaches, which are both biological and non-biological inspired. The different parts are discussed in detail:

In the *security environment* run the different common used protection components as well as the artificial cells. The security environment manages the access to the resources of the node. It provides a common interface for this access like a middleware between the node and the protection components, which work platform-independent. Furthermore, the security environment provides only the allowed components access to resources. For this is a distributed public key infrastructure used, where each component authorizes during the access to the resources. The security environment is so installed that an adversarial cannot use it for attacks. Therefore, the security environment is encrypted installed and the communication between the security environments is also encrypted. In SANA, the approach is to install the security environment as a virtual machine using hardware virtualization - e.g. OpenVZ or KVM. A second virtual machine contains the operating system for the user. The underlying operating system is only used to utilize the virtual environments and is only changed when the hardware is changed. The security environment is allowed to check the underlying operating system and all other virtual machines where other virtual machines are not allowed to see other virtual machines. Then, an intrusion in the underlying operating system can be detected because this system changes only when the hardware changes. An intrusion in the normal operating system cannot see the security environment and this is also secured against attacks through integrity checks. The different layers in the implementation are also visualized in figure 1.

The protection components consist of two types, which are installed in the security environment. The common used protection components like antivirus softwares, firewalls, packet filters, and intrusion detection systems are installed and registered in a security environment without any changes to the internal workflows of the components. An artificial cell connects a common used security component to the security environment in order to translate the warnings and alerts, to inform the component about events, and to check and update the security component. The second type is the population of *artificial cells* [14]. Each cell is a lightweighted and mobile agent performing certain tasks in the security environments and moving through the network. The tasks are various: e.g. checking the packets, files, or system calls for intrusions [13], identifying infected nodes [12], performing regular checks, and collecting status information. However, the cells can perform all tasks and novel techniques are deployed through novel artificial cells. E.g., the packet checking cells build up a distributed intrusion detection system protecting all nodes against intrusions packed in network packets. The warnings and alerts generated by the protection components are combined in a common log file for each security environment. The information collected in the system are used to learn the system, e.g. to develop new strategies to defend the network - self-learning.

SANA uses lots of different artificial cells, which are redundant installed in the network. The workflows consist of several small tasks where each task is performed by one cell. For the collaboration is the *artificial cell communication* used, which is a robust, fault-tolerant, and efficient communication protocol for point to multi-point communication. Two specialized nodes are added: the artificial lymph nodes supply the protection components of a small area with additional information and respond to important messages. Furthermore, the artificial lymph nodes decide when a message is sent to all nodes of the network and they collect information about the supplied network part. The second specialized node is the central nativity and training station (CNTS) that implements an artificial bone marrow and thymus. It generates and releases continuously novel artificial cells in order to keep the population of cells up-to-date. Thereby, it also includes novel approaches of network security like enhancements and updates. In addition, it collects status information for the administrator and for further analysis.

The various artificial cells move through the network autonomously. However, it must be always guaranteed that each node is properly secured. Therefore is the *self-management* used where each security component knows the re-quired security it provides - the security value. Each node calculates its security level basing on the security values and when this level falls below a certain threshold, it starts a notification process. This process attracts cells from nearby nodes to move to this node and that cells in this node do not leave; however, the cells still work autonomously. An additional workflow of the self-management organizes the cells, which flow through the network and perform certain tasks, so that each node is regularly checked but not too often to save resources.

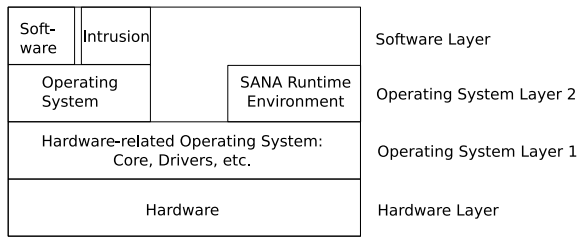

Figure 1: Implementation layers in a network node. The operating system and the SANA runtime environment in operating system layer 2 run in different hardware virtualizations. The operating system in layer 1 changes only when the hardware changes.

Different protection components work in the artificial immune system SANA and each component performs other tasks as well as gains other information. This information should be exchanged in order to identify abnormal behavior quickly. Therefore, the artificial cell communication is used to exchange messages between a sender and a set of receivers in a small area of the network. Important messages are also broadcasted to all receivers in the network or immediately to the administrator. Furthermore, the cells exchange every time step summary information about the current status in order to inform the nearby cells. All of these cells calculate a danger level and adapt their internal behavior accordingly - implementation of the *danger theory*. Other adaptive workflows are to identify also modified intrusions using a similarity measurement and that the cells move to the area where the attacks are more likely to occur. SANA also identifies infected nodes using a special type of artificial cells in reusing the information from components checking network packets. Then, the system quarantines these nodes and starts a disinfection process.

The maintenance workflows are different to common used security systems. The artificial cells shutdown over time and the CNTS release novel cells over time. These novel cells include the newest information and techniques about intrusions and the population is always up-to-date. The other components are updated using updating cells, which flow through the network and report each update; thus, unsuccessful updates and outdated components are identified. Checking cells furthermore test the protection components and report not proper working components. These cells also check if a node is identified with a certain intrusion. Novel approaches of network security are introduced using a novel type of artificial cells and are quickly released by the CNTS because they work platform-independent in the security environment. The administrator can always demand a status snapshot using the information collected by the CNTS and they can access each security environment using the network for more detailed information. The administrator can define the granularity of the demanded information and the system collects the information using artificial cells and the information stored in the artificial lymph nodes and CNTS. Important messages are immediately sent to the administrator using the artificial cell communication. The maintenance workflows are implemented through a management software, which enables access to all security environments and their security components. The administrator uses this management software to administrate SANA and its components and to observe the working of the nodes.

Due to the virtualization of SANA's security environment and the operating system, they are no longer platform dependent. Consequently, the system is able to duplicate, halt, and transfer them over the network. This leads to a complete different view of the system where a service oriented architecture is introduced. The artificial lymph nodes as well as the CNTS provide on demand security components as virtual machines, which are transferred to the demanding security environment when a certain event occurs to perform additional non-standard processes. Additionally, as infected identified operating systems are duplicated in order to save the legal evidences and to move these to CNTS for further analysis to enhance the system. In addition, the operating systems can be organized as services so that an user works always with the same system and can also quickly move the operating system to other nodes. However, the latter issue is not in the scope of the SANA system.

SANA's project status is that it is implemented on a network simulator as a proof-of-concept implementation. The automatic workflows work well and the adaptive behavior increases the performance. Using the different types of artificial cells are redundancies installed so that a partly breakdown does not influence the overall system. The results of SANA are explained in the next section.

## 6. RESULTS

The performance of SANA is more than acceptable. Using the maintenance workflows, the system can be easily updated and extended. The system processes the warnings and alerts automatically and uses this information in order to configure and adapt itself so that it copes with the current situation. Important information are quickly delivered to the administrator, who can influence the system. The status information are collected in the CNTS and the administrator can always access the information e.g. for further analysis. The self-management organizes the various artificial cells so that all nodes are properly secured and regular checks are performed on all nodes. The artificial cell communication works well so that the artificial cells can quickly and reliably exchange messages for collaboration.

AGNOSCO - a special type of artificial cells identifying infected nodes using the information gathered by components analyzing the network traffic [12] - helps to keep the system free of infections and improves the performance significantly because common used systems mostly do not identify infected nodes. Furthermore, AGNOSCO uses the information from lots of protection components, which are distributed over nearly all nodes. Other cells perform regular checks in order to identify infections, not proper working components, and abnormal behavior - self-checking. These cells report infections quickly and increase the performance because a not proper working protection components are a risk for the whole network. After identifying a problem, the system develops a strategy to solve it - self-repairing and -healing. Examples are to add novel artificial cells for the identification of new intrusions or to disinfect a node using an artificial cell. Otherwise, the system quarantines the node quickly and informs the administrator for disinfection.

Most attacks towards networks are not only performed towards a single node but rather towards several nodes in order to find a weak point in the network - e.g. IP-range scanning; to cope with this is the artificial clonal selection and hypermutation used. In order to find such attacks, SANA combines the information gathered in different nodes for the identification of abnormal attacks through a more sophisticated information management. One example is the already explained AGNOSCO approach. Other approaches are to combine summaries of the logs of different nodes and analyze these in order to find patterns describing abnormal behavior. For cooperation and adaptively, the artificial cells exchange continuously summary status information in order to adapt their internal thresholds so that the overall system SANA adapts quickly to the current situation, which is a first implementation of the danger theory and will be extended in the near future.

The distinguished installation through hardware virtualization enables several new features where especially the service oriented architecture (SOA) helps to demand the required components in order to react to the certain events like identification of infections. The duplication of virtual machines works well and the intrusions can be stored in order to analyze them in detail and to use the legal evidences for legal analysis. The distinguished installation furthermore secures the security system against attacks because the normal operating system is not able to attack the security environment and the security environment is secured with integrity checks.

For the evaluation of SANA, different attack scenarios are implemented and simulated. SANA performs well and protects the network efficiently:

**Worm, Virus, and Trojan Attacks** A worm attacks the node over the network using some security holes. The worm mostly installs itself in the node - infection called - and performs certain tasks; the virus is installed when the host-programm is executed. Often, they open a backdoor so that other intrusions or hackers can occupy the node. SANA detects and removes the network packets containing the worm or virus, when it is known by SANA. Infected nodes are identified and disinfected as well as backdoors are identified by regular checks; the information are used in order to develop a protection strategy. Thus, SANA protects the network against worms as well as viruses and also disinfects it. Due to the regular checks is the performance higher than in common used protection systems. With the combined analysis of different nodes and the status messages sent every time step are multi-step or multi-stage attacks identified because each step is detected and reported to the nearby cells.

**Encrypted Traffic Attacks** These attacks install a backdoor in some node and connect to this node over the network with encrypted traffic. The network packets can be only checked at the source and destination node of the connection where the source node is the hacker's node and normally not accessible. SANA uses a distributed IDS installed by the artificial cells and checks the network traffic on all nodes. Thus, infected packets are identified and removed at the destination node and the backdoor is identified as well as removed by regular checks performed by artificial cells.

**Hacker Attacks** The identification of hacker attacks is more difficult than identifying a worm or virus attack. For preventing a hacker attack, SANA uses at least the same workflows as current protection systems. Additionally artificial cells check regularly the nodes for installed backdoors - e.g. VPNserver with IPsec -, which are used by the hacker for further entries to the node. Also, other changes and additionally installed software is detected and reported to the administrator in order to secure the network more properly.

**Social Attacks** This type of attack is performed by users with internal information, e.g. an unsatisfied employee trying to gain internal information for further usage. The problem is to distinguish between the normal and abnormal network usage of such an user. However, some characteristics are e.g. that the user tries to access resources, which are normally not used by this user, and the user tries different passwords generating login errors. Analyzing only the information from a single node is mostly not sufficient but combining the information from several nodes gives a hint in order to identify such an user. Therefore, SANA uses artificial cells analyzing the log files of all nodes and identifies these users.

**Physical Attacks** The adversarial has physical access to the node. The adversarial can connect external storages, additional hardware, or start another operating system. SANA identifies changes in the security environment and the hardware virtualizaton system through integrity checks and does not start the infected system. Furthermore, the neighbor nodes of a starting machine isolate the node until the security environments of the nodes synchronize and enable the network access. Without a running security environment is a network access not possible and the neighbor nodes report the nodes without proper security environment. Thus, physical attacks and changes in the security systems are detected and prevented.

To sum up, SANA shows that a distributed approach like an artificial immune system enables additional features not provided by common used protection systems. These features are e.g. analysis combining information from lots of different nodes, checking from outside through the installation of the security environment, and the cooperation between the numerous small components. Furthermore, the fast deployment of extensions and updates as well as the automatic workflows without interaction with the user. The virtualization enables further features like the duplication of infected virtual machines. With theses features, the performance of SANA is significantly better compared to common used protection systems because SANA also adapts to the current situation in the network as well as SANA can be quickly enhanced with novel approaches..

## 7. CONCLUSION

Artificial immune systems provide several advantages in protecting a network against intrusions compared to common used protection systems. Especially the architecture using lots of lightweighted, mobile, and autonomous artificial cells without a centralized server facilitates a more sophisticated information management and a self-checking for the identification of abnormal behavior. The checking from outside as well as the quickly deployment is important for coping with upcoming intrusions. In order to bring an artificial immune system into production, several tasks have to be done where especially the implementation and the design are challenges. SANA is a framework of an artificial immune system and shows that the performance of artificial immune systems in the network security domain is more than acceptable. The next steps are to simulate more realistic attack scenarios and to discuss how to implement the security environment in each node.